\documentclass{article}
\usepackage{spconf,amsmath,graphicx}
\usepackage{hyperref}
\hypersetup{
    colorlinks=true,
    linkcolor=black,
    filecolor=black,      
    urlcolor=black,
    citecolor=black,
    pdftitle={Overleaf Example},
    pdfpagemode=FullScreen,
    }

\title{Chord-Conditioned Melody Harmonization \\ with Controllable Harmonicity}
\name{Shangda Wu$^{\star}$ \qquad Xiaobing Li$^{\star}$ \qquad Maosong Sun$^{\star \dagger}$}
\address{$^{\star}$ Department of Music AI and Information Technology, Central Conservatory of Music \\
$^{\dagger}$Department of Computer Science and Technology, Tsinghua University}
\begin{document}
%
\maketitle
\begin{abstract}
Melody harmonization has long been closely associated with chorales composed by Johann Sebastian Bach. Previous works rarely emphasised chorale generation conditioned on chord progressions, and there has been a lack of focus on assistive compositional tools. In this paper, we first designed a music representation that encoded chord symbols for chord conditioning, and then proposed DeepChoir\footnote{\href{https://github.com/sander-wood/deepchoir}{https://github.com/sander-wood/deepchoir}}, a melody harmonization system that can generate a four-part chorale for a given melody conditioned on a chord progression. With controllable harmonicity, users can control the extent of harmonicity for generated chorales. Experimental results reveal the effectiveness of the music representation and the controllability of DeepChoir.
\end{abstract}
\section{Introduction}\label{sec:introduction}
Chorales are vocal or instrumental music performed by a group of people, and the four-part setting for SATB (Soprano, Alto, Tenor, and Bass) is the standard for chorales. It is necessarily ``polyphonic", i.e., consisting of two or more voices. It takes a lot of practice to establish a relationship between voices that is harmonically and rhythmically interdependent.

The task of chord-conditioned melody harmonization is aimed at \textbf{automatically composing a four-part chorale based on a melody and a chord progression}. Formally, given the soprano (melody) sequence, $S_{1:t} = \{S_{1}, S_{2}, ... , S_{t}\}$ and the chord sequence $C_{1:t} = \{C_{1}, C_{2}, ... , C_{t}\}$, the goal is to infer sequences of the other three voices, i.e., $A_{1:t}$, $T_{1:t}$ and $B_{1:t}$. All sequences (except for $C_{1:t}$) share the same vocabulary $V$ of size $|V|\approx100$. The phrase \textit{controllable harmonicity} refers to \textbf{the ability for users to encourage systems to generate more (or less) chord tones}.

\begin{figure}[t]
	\centering
		\begin{minipage}{6.2cm} 
            \includegraphics[width=\textwidth]{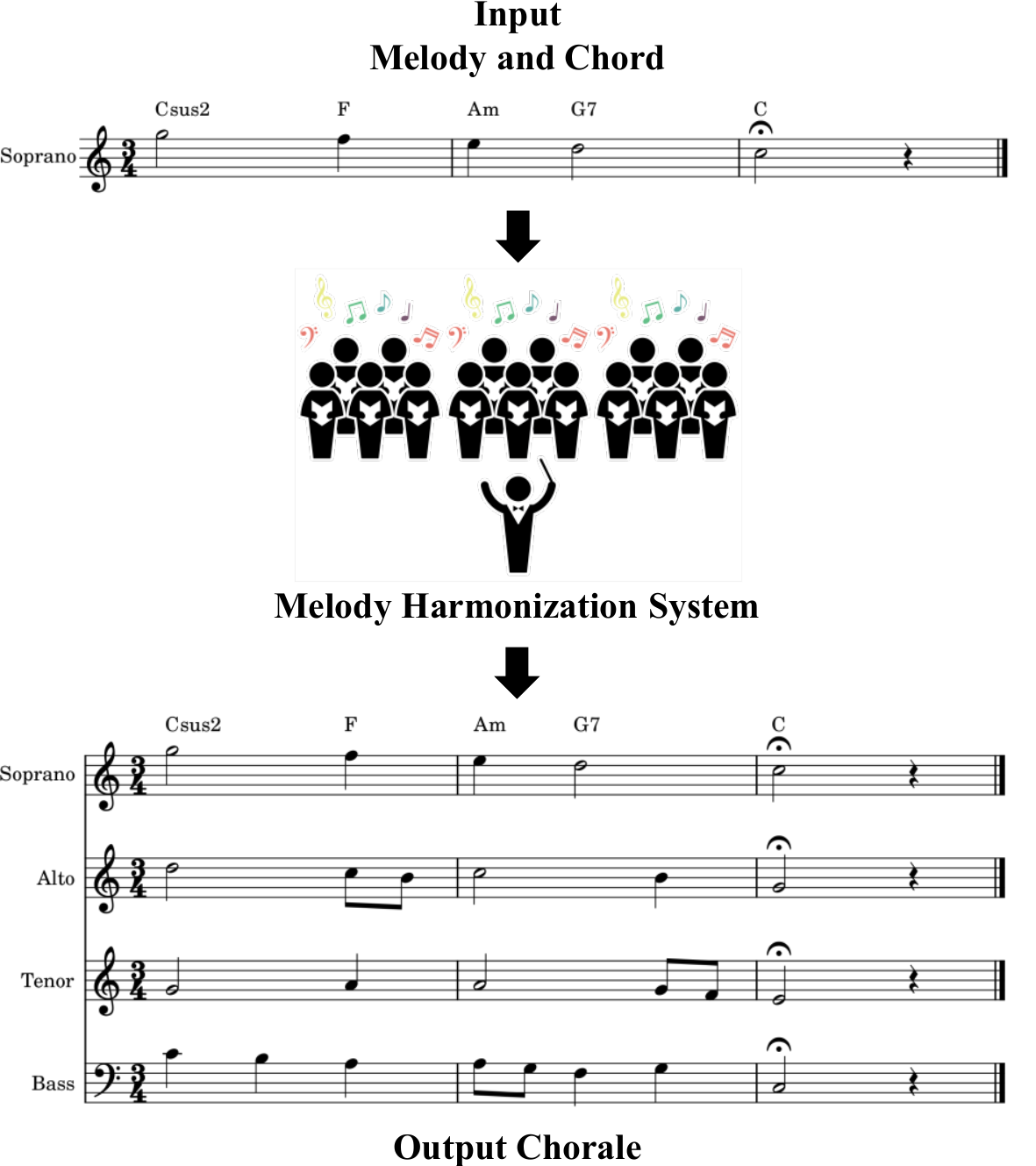}
		\end{minipage}
    \centering
	\caption{The process of melody harmonization. Here, the alto, tenor, and bass are generated by the system.} 
	\vspace{-1em}
\end{figure}

Studies of melody harmonization \cite{DBLP:journals/corr/abs-2110-02640,DBLP:conf/pkdd/LeemhuisWH19,DBLP:conf/icmc/Ebcioglu86,DBLP:journals/corr/abs-1907-06637} have long focused on the Johann Sebastian Bach (JSB) Chorales dataset, and their goal is to make the music generated by systems sound as much like Bach as possible. For example, BachBot-generated chorales \cite{DBLP:conf/ismir/LiangG0S17} are almost indistinguishable from those of Bach, according to their subjective experiment results. Meanwhile, DeepBach \cite{DBLP:conf/icml/HadjeresPN17}, a system aimed at modelling polyphonic music, has been successfully applied to Monteverdi's five-part madrigals and Palestrina's masses, even if this system mainly focuses on JSB Chorales. As for non-neural network solutions like BacHMMachine \cite{DBLP:journals/corr/abs-2109-07623}, a probabilistic framework that integrates compositional principles, is capable of generating musically convincing chorales.

Although the above-mentioned works reproduce the style of Bach quite well, they \textbf{lack chord conditioning} (the learning of chord transitions), and most existing melody harmonization systems are \textbf{uncontrollable}.

In this paper, our goal is to provide an assistive compositional tool that can help users easily compose chorales. We first automatically labelled chord symbols for JSB Chorales, and encoded chords as chromagrams. We then added beat information into our music representation to make the system aware of time signatures. Furthermore, we applied gamma sampling \cite{https://doi.org/10.48550/arxiv.2205.06036}, a sampling method for controlling language models, to achieve controllable harmonicity in the melody harmonization task, enabling users to steer the extent to which the system follows a given chord progression.

\begin{figure}[t]
	\centering
		\begin{minipage}{8.9cm} 
            \includegraphics[width=\textwidth]{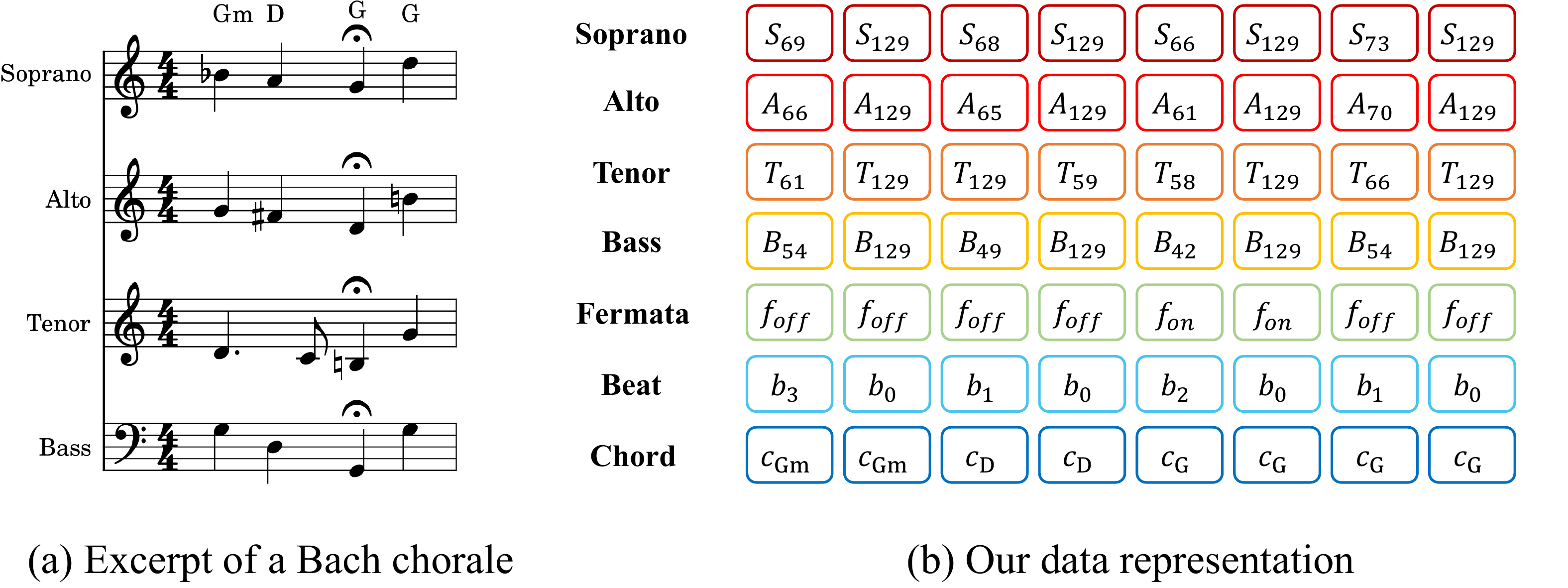}
		\end{minipage}
    \centering
    \vspace{-0.5em}
	\caption{For demonstration purposes, the time resolution of our proposed music representation is set to eighth notes.} 
	\vspace{-0.5em}
\end{figure}

\section{Methodology}
\subsection{Music Representation}

As shown in Fig. 2b, our symbolic representation encodes each chorale into the following 7 aligned sequences.

\begin{itemize}
    \item 
    \textbf{SATB Sequences}: encode each voice as a piano-roll consisting of 130-D one-hot vectors (128 pitches, 1 rest, and 1 hold), with a sixteenth note time resolution.
    
    \item 
    \textbf{Fermata Sequence}: consists of boolean values indicating whether a fermata symbol is present in each frame.
    
    \item 
    \textbf{Beat Sequence}: based on time signatures, encodes the beat information into 4-D one-hot vectors, which correspond to non-beat, weak, medium-weight, and strong beats (ranging from 0 to 3).
    
    \item
    \textbf{Chord Sequence}: encodes chords as chromagrams (12-D multi-hot vectors), and each dimension corresponds to an activated pitch class of a chord.
\end{itemize}

\begin{figure}[t]
	\centering
		\begin{minipage}{7.5cm} 
            \includegraphics[width=\textwidth]{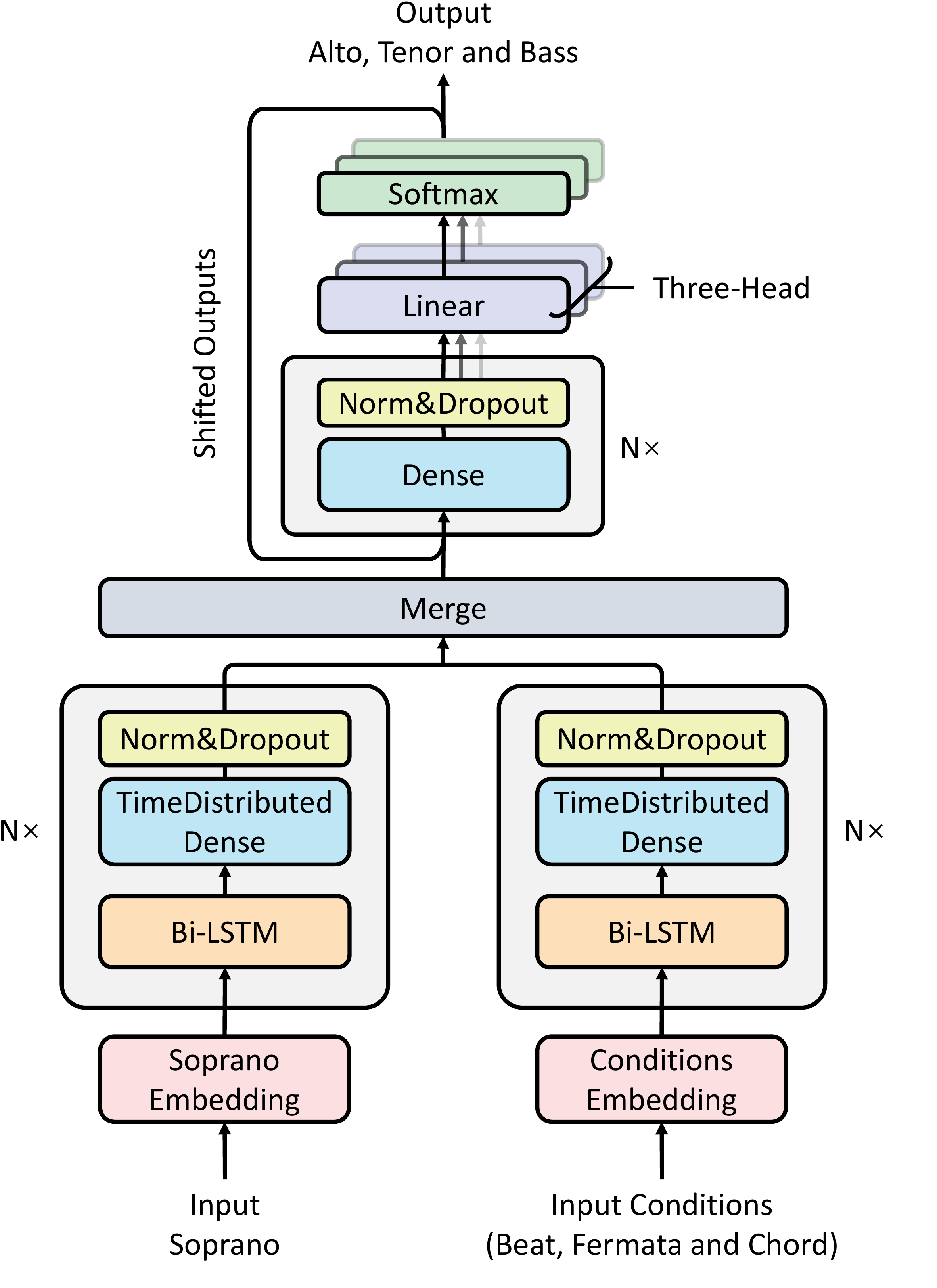}
		\end{minipage}
    \centering
	\caption{The architecture of DeepChoir.} 
	\vspace{-1em}
\end{figure}

\subsection{DeepChoir}
DeepChoir is a \textbf{multi-headed autoregressive model with an encoder-decoder structure}, as shown in Fig. 3. The two encoders use a stacked Bi-LSTM \cite{DBLP:journals/neco/HochreiterS97,DBLP:journals/tsp/SchusterP97} with a time-distributed dense layer, and the decoder uses a stacked dense layer.

\textbf{Encoder. }The structure of these two encoders is identical, but their weights are not shared. The input of the soprano encoder is the embedded soprano sequence, whereas the conditions encoder one is a concatenation of embedded beat, fermata, and chord sequences. The two encoders are formed by a stack of $N=3$ identical blocks with two sub-layers. The first is a 128-hidden-size Bi-LSTM layer, and the second is a time-distributed dense layer. After the time-distributed dense layer, we use batch normalization \cite{DBLP:conf/icml/IoffeS15} followed by dropout \cite{DBLP:journals/corr/abs-1207-0580} with a rate of 0.2.

\textbf{Decoder. }The decoder, like the other two encoders, is composed of a stack of $N=3$ identical blocks, but it is much simpler, including only a naive stacked dense layer with batch normalization and dropout. The three-head linearly transforms and softmaxes the output of the decoder, and autoregressively generates the ATB voices frame-by-frame.

\subsection{Controllable Harmonicity}

Based on the assumption that some attributes of the generated text are closely related to the occurrences of certain tokens, gamma sampling \cite{https://doi.org/10.48550/arxiv.2205.06036} achieves controllable text generation via \textbf{scaling probabilities of attribute-related tokens during generation time}:

\begin{equation}
\resizebox{.85\hsize}{!}{$
\begin{aligned}
    p_{\mathcal{A}_{out}}&=p_{\mathcal{A}_{in}}^{tan(\frac{\pi \Gamma}{2})}, \\
    p_{a_{out}}&=p_{a_{in}}\cdot \frac{p_{\mathcal{A}_{out}}}{p_{\mathcal{A}_{in}}},\quad \forall a\in \mathcal{A}, \\
    p_{n_{out}}&= p_{n_{in}} \cdot (1 + \frac{p_{\mathcal{A}_{in}}-p_{\mathcal{A}_{out}}}{p_{\backslash \mathcal{A}_{in}}}),\quad \forall n\notin \mathcal{A},\\
\end{aligned}$}
\label{eq3}
\end{equation}

\noindent
where $\Gamma\in[0,1]$ is the user-controllable control strength, $\mathcal{A}$ is the set of attribute-related tokens (${\backslash \mathcal{A}}$ is its complement), $p_{a_{in/out}}$ is the input/out probability of an attribute-related token $a$, and the same goes for every non-attribute-related token $n$. When $\Gamma=0.5$, there is \textbf{no change in the probability distribution}, while when $\Gamma<0.5$, \textbf{the probabilities of the attribute-related tokens increase} and vice versa.

Since harmonicity depends on the proportion of chord tones present in the music (the larger the proportion, the more harmonious it is), we define chord tones as the attribute-related tokens for controllable harmonicity. To make the controllable harmonicity more intuitive, we set the harmonicity parameter $h=1-\Gamma$.

\begin{figure}[t]
	\centering
		\begin{minipage}{8cm} 
            \includegraphics[width=\textwidth]{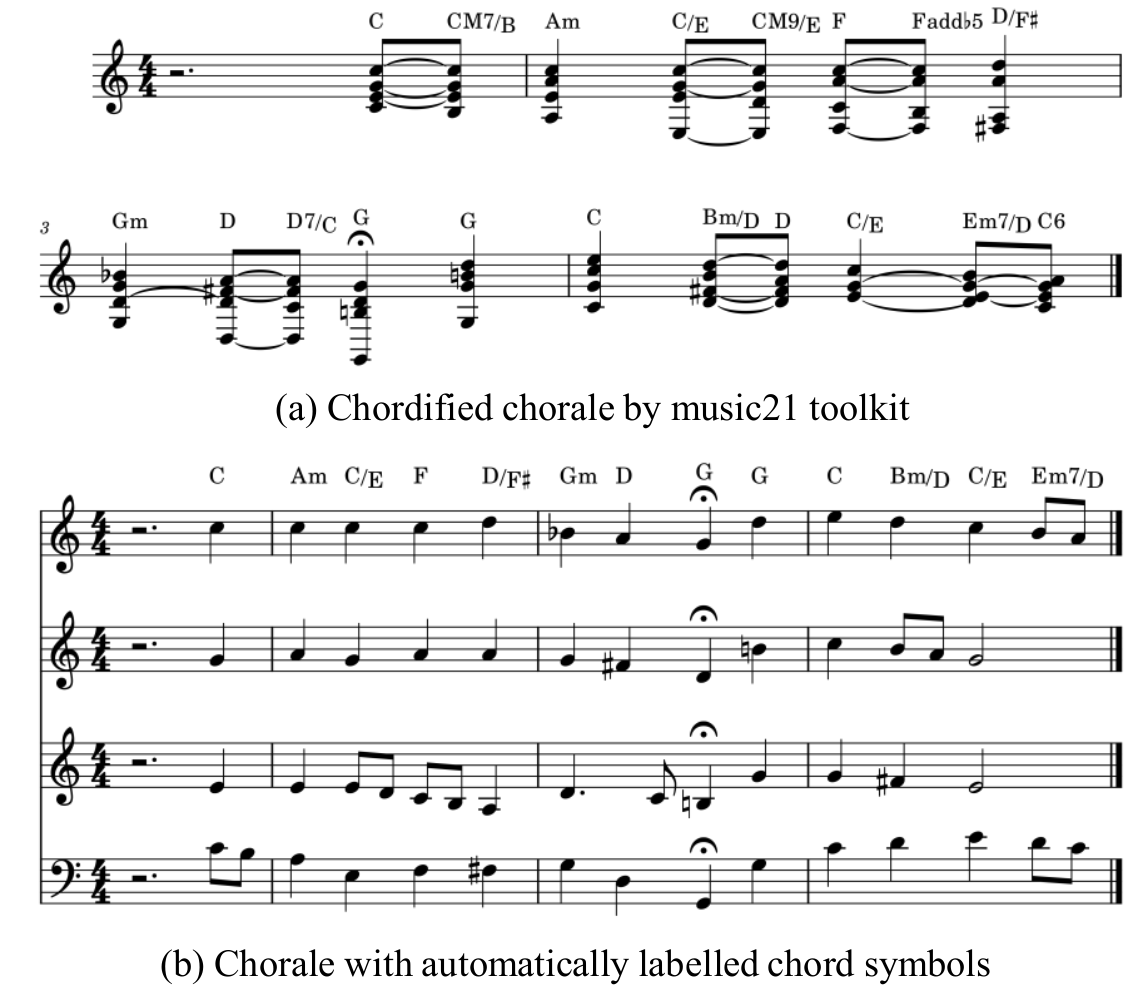}
		\end{minipage}
    \centering
	\caption{An excerpt of the processed chorale (BWV 347). The chord symbols in Fig. 4b are converted from Fig. 4a.} 
\end{figure}

\section{Experiments}
\subsection{Dataset}
As we did not find any full JSB Chorales dataset with human-annotated chord symbols, and as performing harmonic analysis manually is too time-consuming, we carried out the following automated preprocessing.

\begin{itemize}
    
    \item 
    \textbf{Chordifying}: simplify a complex score with multiple voices into a succession of chords in one voice (see Fig. 4a) via a tool in music21 \cite{DBLP:conf/ismir/CuthbertA10}.
    
    \item 
    \textbf{Labelling}: rank all chords based on beat strength, then only keep those with the highest scores and add them to sopranos (see Fig. 4b).
\end{itemize}

We ended up with a total of 366 chorales for training (90\%) and validation (10\%). Admittedly, these automatically labelled chord symbols are not authoritative from the perspective of harmonic analysis, but they are sufficient for our task.

\subsection{Evaluations of Music Representation}
\subsubsection{Baselines}
In addition to the proposed DeepChoir system, we conducted ablation studies to validate the importance of chord progression and beat information. For all systems below, $h = 0.5$.

\begin{itemize}
    \item 
    \textbf{DeepChoir (w/o chord)}: based on DeepChoir (full), the chord sequences are removed.
    \item 
    \textbf{DeepChoir (w/o beat)}: based on DeepChoir (full), the beat sequences are removed.
\end{itemize}

Representations of two well-known melody harmonization systems \cite{DBLP:conf/ismir/LiangG0S17, DBLP:conf/icml/HadjeresPN17} were chosen as baselines. It should note that their task (i.e., plain melody harmonization) is not exactly the same as DeepChoir. To the best of our knowledge, there are no other chord-conditioned melody harmonization systems available. Our purpose in choosing them as baselines is to evaluate the performance gains of introducing beat information without chords.

\begin{itemize}
    \item 
    \textbf{BachBot} \cite{DBLP:conf/ismir/LiangG0S17}: only considers fermata information as the condition (no beat and chord sequences).
    
    \item 
    \textbf{DeepBach} \cite{DBLP:conf/icml/HadjeresPN17}: uses a subdivision list which includes indexes of sub-beats without considering time signatures (with fermata and beat-like sequences but no chord sequences). 
\end{itemize}

\subsubsection{Analyses of Music Representation}

\begin{figure}[t]
	\centering
		\begin{minipage}{8cm} 
            \includegraphics[width=\textwidth]{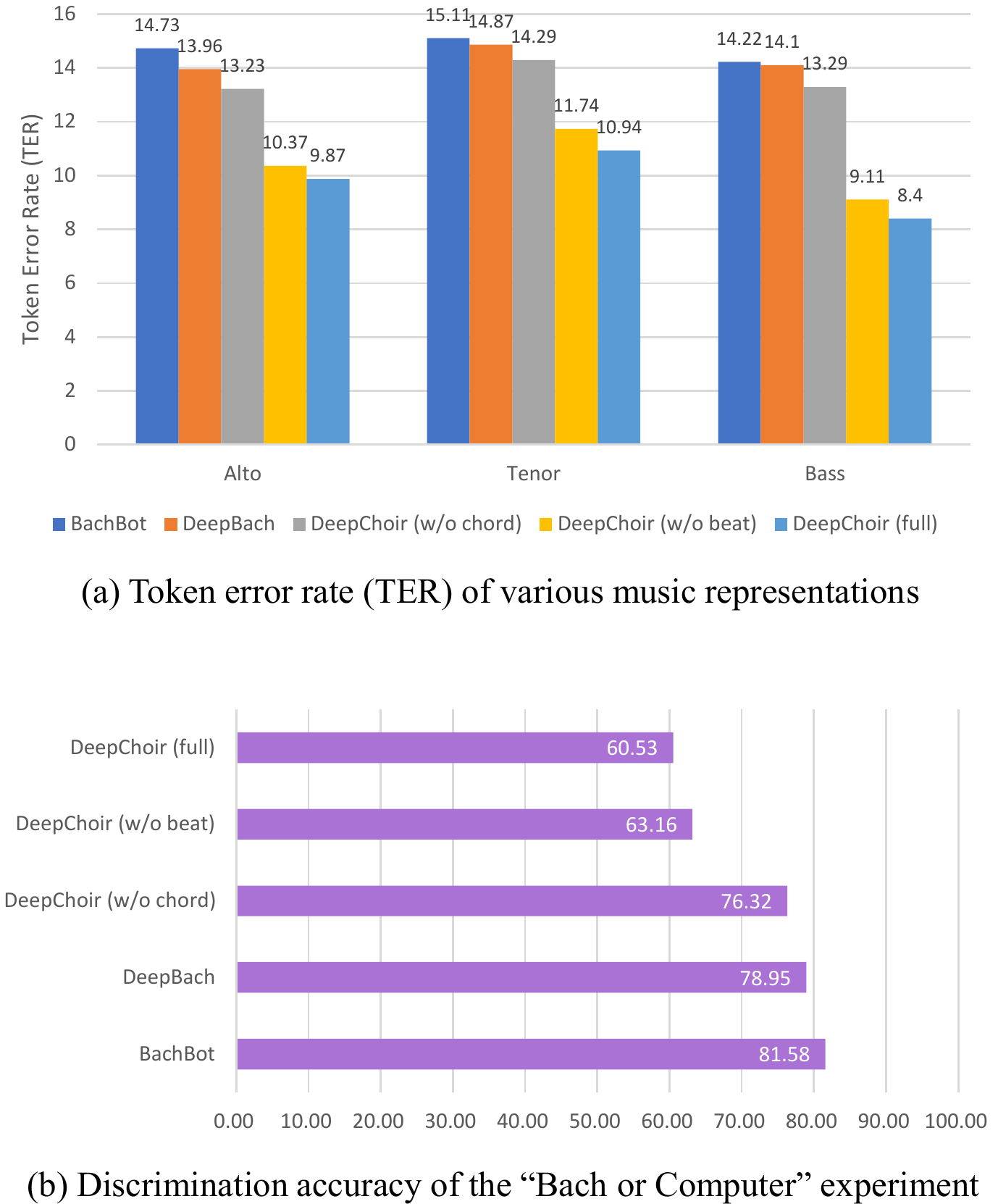}
		\end{minipage}
    \centering
	\caption{Results of various music representations.} 
	\vspace{-1em}
\end{figure}

The validation set (38 chorales) reharmonized by different systems was used for this evaluation. We chose the token error rate \cite{DBLP:conf/ismir/LiangG0S17} (TER, \% of errors in token prediction) for quantitative evaluation. For the qualitative evaluation, we invited students from music schools to take a discrimination test, and received a total of 190 feedbacks.

As shown in Fig. 5a, DeepChoir (w/o chord) offers a moderate improvement (about 1\%) over DeepBach and BachBot in terms of TER, which suggests the effectiveness of encoded beat information. After introducing the chord information, both DeepChoir (w/o beat) and DeepChoir (full) show a significant decrease (about 4\%) in TER compared to DeepChoir (w/o chord). This demonstrates that the generation constrained by chords is more consistent with ground truth at the token level. In addition, the improvement is more noticeable on bass, as once chords are given, bass notes can be more deterministic, compared to alto and tenor.

The results of the discrimination test (see Fig. 5b) are quite clear: as music representation complexity rises, discrimination accuracy falls. When presented with a chorale generated by DeepChoir (full), about 40\% of the listeners believed it was composed by Bach. Given the complexity of Bach's compositions and the fact that some of the listeners are already familiar with JSB Chorales, we consider this to be an excellent performance.

\subsection{Evaluations of Controllability of Harmonicity}

\subsubsection{Metrics}
Objective evaluations used three chord/melody harmonicity metrics proposed in \cite{DBLP:journals/corr/abs-2001-02360}, which have been extensively used in the melody harmonization task \cite{DBLP:conf/icassp/SunCLCW21,DBLP:journals/corr/abs-2108-00378,DBLP:journals/access/RhyuCKL22}.

\begin{itemize}
    \item 
    \textbf{Chord Tone to non-Chord Tone Ratio (CTnCTR)}: calculates the ratio of the number of the chord tones to the number of the non-chord tones.
    
    \item 
    \textbf{Pitch Consonance Score (PCS)}: calculated based on the interval between melody and chord notes.
    
    \item 
    \textbf{Melody-Chord Tonal Distance (MCTD)}: represents a melody note by a pitch class profile feature vector \cite{DBLP:conf/icmc/Fujishima99} and compares it against a chord symbol in the 6-D tonal space \cite{harte2006detecting} to calculate the closeness between them.
\end{itemize}

For subjective evaluations, we invited students from music schools to rate all generated chorales from 1 to 5 based on their quality, and a total of 266 ratings were received. For presentation purposes, the scores were rescaled to 0.2 to 1 as a subjective metric, i.e., the \textbf{Rescaled Score (RS)}.

\begin{figure}[t]
	\centering
		\begin{minipage}{7.9cm} 
            \includegraphics[width=\textwidth]{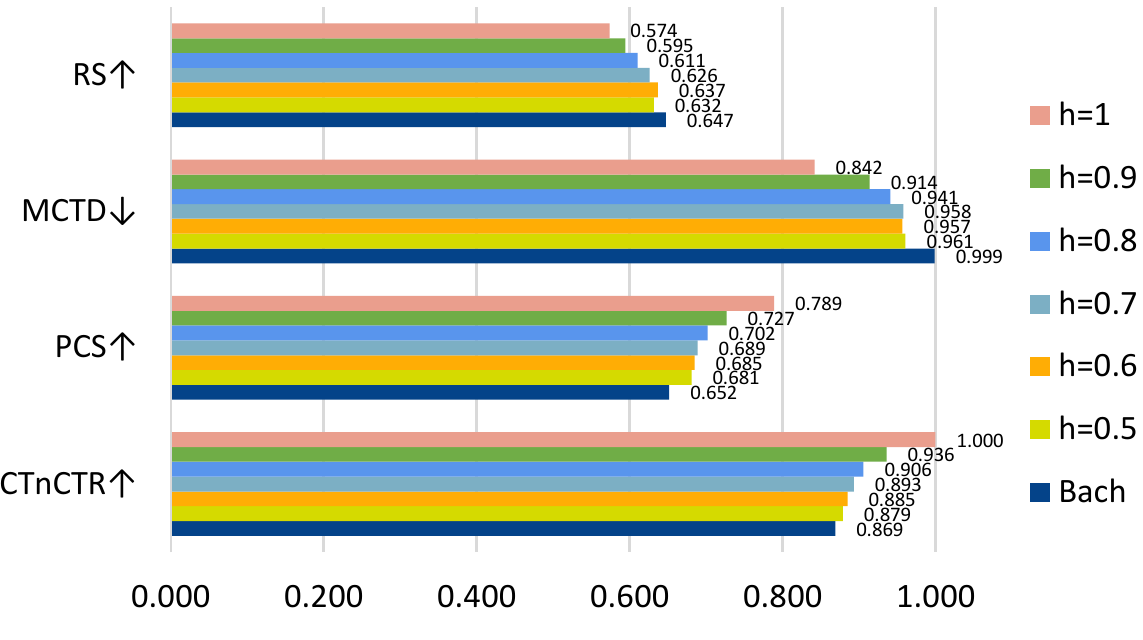}
		\end{minipage}
    \centering
    \vspace{-0.5em}
	\caption{Results of various harmonicity settings.}
	\vspace{-1em}
\end{figure}

\subsubsection{Analyses of Controllable Harmonicity}
We reharmonized the validation set (38 chorales) with $h$ ranging from 0.5 (original) to 1.0 (blocking all non-chord tones).

As shown in Fig. 6, $h$ has strong positive correlations with the CTnCTR and PCS, with Pearson correlation coefficients of $r = 0.9 $ and $r = 0.88 $, respectively, but an obvious negative correlation with MCTD, where $r = -0.86 $. These indicate that gamma sampling can significantly tune the harmonicity of generated chorales. However, according to RS, when $h>0.6$, the listener's rating of the generated chorales decreases as $h$ increases. We suggest that this is because setting $h=0.6$ avoids the generation of some inappropriate non-chord tones, but further increasing $h$ leads to the chorales becoming more bland and inconsistent with Bach's style.

\begin{figure}[t]
	\centering
		\begin{minipage}{7.75cm} 
            \includegraphics[width=\textwidth]{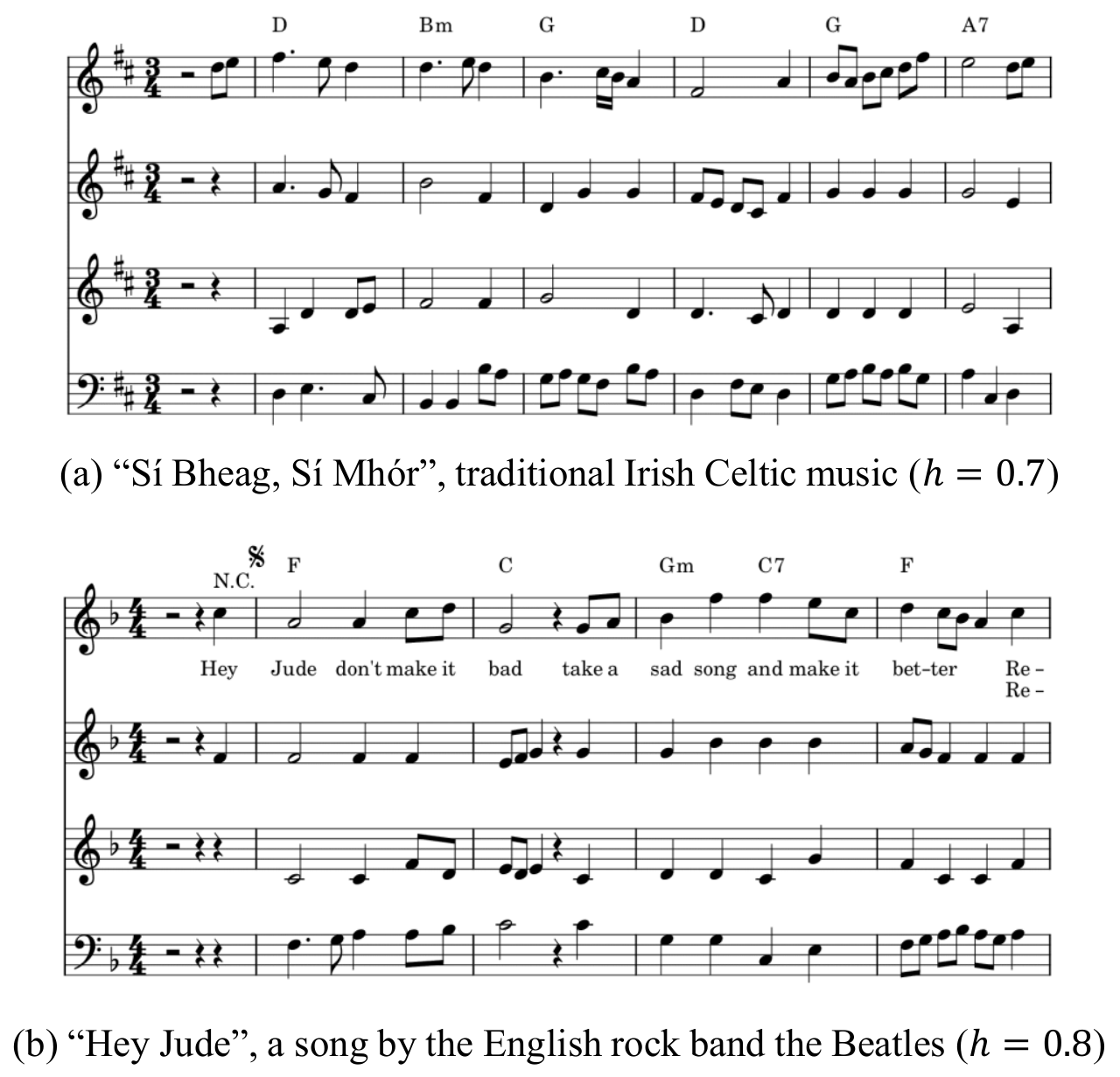}
		\end{minipage}
    \centering
	\caption{DeepChoir-generated examples based on music with completely different styles from the JSB Chorales.}
	\vspace{-1em}
\end{figure}

\section{Generation Examples}
Fig. 7 provides examples generated by DeepChoir. We empirically tuned $h$ in keeping with the original styles. Despite melodies and chords being taken from completely different styles compared to the training data, the generated chorales are still musically convincing. The expressive use of non-chord tones, especially in the bass, is quite impressive. With conditioned chords, these chorales not only match their original styles, but harmonically fit into existing accompaniments.

\section{Conclusions}
We proposed DeepChoir, a chord-conditioned melody harmonization system with controllable harmonicity. The experiment results demonstrate that DeepChoir-generated chorales are more consistent with ground truth than baselines, and tuning the controlling parameter $h$ can significantly change their harmonicity. We expect that, with chord conditioning and controllable harmonicity, this system can be used to compose general polyphonic music that is not limited to the Bach style.

\bibliographystyle{IEEEbib}
\bibliography{strings,refs}

\begin{thebibliography}{10}

\bibitem{DBLP:journals/corr/abs-2110-02640}
Minghe Kong and Lican Huang,
\newblock ``Bach style music authoring system based on deep learning,''
\newblock {\em CoRR}, vol. abs/2110.02640, 2021.

\bibitem{DBLP:conf/pkdd/LeemhuisWH19}
Alexander Leemhuis, Simon Waloschek, and Aristotelis Hadjakos,
\newblock ``Bacher than bach? on musicologically informed ai-based bach chorale
  harmonization,''
\newblock in {\em Machine Learning and Knowledge Discovery in Databases -
  International Workshops of {ECML} {PKDD} 2019, W{\"{u}}rzburg, Germany,
  September 16-20, 2019, Proceedings, Part {II}}. 2019, vol. 1168 of {\em
  Communications in Computer and Information Science}, pp. 462--469, Springer.

\bibitem{DBLP:conf/icmc/Ebcioglu86}
Kemal Ebcioglu,
\newblock ``An expert system for harmonizing four-part chorales,''
\newblock in {\em {ICMC} 1986, Den Haag, The Netherlands, October 20-24, 1986}.
  1986, Michigan Publishing.

\bibitem{DBLP:journals/corr/abs-1907-06637}
Cheng{-}Zhi~Anna Huang, Curtis Hawthorne, Adam Roberts, Monica Dinculescu,
  James Wexler, Leon Hong, and Jacob Howcroft,
\newblock ``The bach doodle: Approachable music composition with machine
  learning at scale,''
\newblock {\em CoRR}, vol. abs/1907.06637, 2019.

\bibitem{DBLP:conf/ismir/LiangG0S17}
Feynman~T. Liang, Mark Gotham, Matthew Johnson, and Jamie Shotton,
\newblock ``Automatic stylistic composition of bach chorales with deep
  {LSTM},''
\newblock in {\em {ISMIR} 2017, Suzhou, China, October 23-27, 2017}, 2017, pp.
  449--456.

\bibitem{DBLP:conf/icml/HadjeresPN17}
Ga{\"{e}}tan Hadjeres, Fran{\c{c}}ois Pachet, and Frank Nielsen,
\newblock ``Deepbach: a steerable model for bach chorales generation,''
\newblock in {\em {ICML} 2017, Sydney, NSW, Australia, 6-11 August 2017}. 2017,
  vol.~70 of {\em Proceedings of Machine Learning Research}, pp. 1362--1371,
  {PMLR}.

\bibitem{DBLP:journals/corr/abs-2109-07623}
Yunyao Zhu, Stephen Hahn, Simon Mak, Yue Jiang, and Cynthia Rudin,
\newblock ``Bachmmachine: An interpretable and scalable model for algorithmic
  harmonization for four-part baroque chorales,''
\newblock {\em CoRR}, vol. abs/2109.07623, 2021.

\bibitem{https://doi.org/10.48550/arxiv.2205.06036}
Shangda Wu and Maosong Sun,
\newblock ``Efficient and training-free control of language generation,''
\newblock {\em CoRR}, vol. abs/2205.06036, 2022.

\bibitem{DBLP:journals/neco/HochreiterS97}
Sepp Hochreiter and J{\"{u}}rgen Schmidhuber,
\newblock ``Long short-term memory,''
\newblock {\em Neural Comput.}, vol. 9, no. 8, pp. 1735--1780, 1997.

\bibitem{DBLP:journals/tsp/SchusterP97}
Mike Schuster and Kuldip~K. Paliwal,
\newblock ``Bidirectional recurrent neural networks,''
\newblock {\em {IEEE} Trans. Signal Process.}, vol. 45, no. 11, pp. 2673--2681,
  1997.

\bibitem{DBLP:conf/icml/IoffeS15}
Sergey Ioffe and Christian Szegedy,
\newblock ``Batch normalization: Accelerating deep network training by reducing
  internal covariate shift,''
\newblock in {\em Proceedings of the 32nd International Conference on Machine
  Learning, {ICML} 2015, Lille, France, 6-11 July 2015}, Francis~R. Bach and
  David~M. Blei, Eds. 2015, vol.~37 of {\em {JMLR} Workshop and Conference
  Proceedings}, pp. 448--456, JMLR.org.

\bibitem{DBLP:journals/corr/abs-1207-0580}
Geoffrey~E. Hinton, Nitish Srivastava, Alex Krizhevsky, Ilya Sutskever, and
  Ruslan Salakhutdinov,
\newblock ``Improving neural networks by preventing co-adaptation of feature
  detectors,''
\newblock {\em CoRR}, vol. abs/1207.0580, 2012.

\bibitem{DBLP:conf/ismir/CuthbertA10}
Michael~Scott Cuthbert and Christopher Ariza,
\newblock ``Music21: {A} toolkit for computer-aided musicology and symbolic
  music data,''
\newblock in {\em {ISMIR} 2010, Utrecht, Netherlands, August 9-13, 2010},
  J.~Stephen Downie and Remco~C. Veltkamp, Eds. 2010, pp. 637--642,
  International Society for Music Information Retrieval.

\bibitem{DBLP:journals/corr/abs-2001-02360}
Yin{-}Cheng Yeh, Wen{-}Yi Hsiao, Satoru Fukayama, Tetsuro Kitahara, Benjamin
  Genchel, Hao{-}Min Liu, Hao{-}Wen Dong, Yian Chen, Terence Leong, and
  Yi{-}Hsuan Yang,
\newblock ``Automatic melody harmonization with triad chords: {A} comparative
  study,''
\newblock {\em CoRR}, vol. abs/2001.02360, 2020.

\bibitem{DBLP:conf/icassp/SunCLCW21}
Chung{-}En Sun, Yi{-}Wei Chen, Hung{-}Shin Lee, Yen{-}Hsing Chen, and
  Hsin{-}Min Wang,
\newblock ``Melody harmonization using orderless nade, chord balancing, and
  blocked gibbs sampling,''
\newblock in {\em {IEEE} International Conference on Acoustics, Speech and
  Signal Processing, {ICASSP} 2021, Toronto, ON, Canada, June 6-11, 2021}.
  2021, pp. 4145--4149, {IEEE}.

\bibitem{DBLP:journals/corr/abs-2108-00378}
Yi{-}Wei Chen, Hung{-}Shin Lee, Yen{-}Hsing Chen, and Hsin{-}Min Wang,
\newblock ``Surprisenet: Melody harmonization conditioning on user-controlled
  surprise contours,''
\newblock {\em CoRR}, vol. abs/2108.00378, 2021.

\bibitem{DBLP:journals/access/RhyuCKL22}
Seungyeon Rhyu, Hyeonseok Choi, Sarah Kim, and Kyogu Lee,
\newblock ``Translating melody to chord: Structured and flexible harmonization
  of melody with transformer,''
\newblock {\em {IEEE} Access}, vol. 10, pp. 28261--28273, 2022.

\bibitem{DBLP:conf/icmc/Fujishima99}
Takuya Fujishima,
\newblock ``Realtime chord recognition of musical sound: a system using common
  lisp music,''
\newblock in {\em Proceedings of the 1999 International Computer Music
  Conference, {ICMC} 1999, Beijing, China, October 22-27, 1999}. 1999, Michigan
  Publishing.

\bibitem{harte2006detecting}
Christopher Harte, Mark Sandler, and Martin Gasser,
\newblock ``Detecting harmonic change in musical audio,''
\newblock in {\em Proceedings of the 1st ACM workshop on Audio and music
  computing multimedia}, 2006, pp. 21--26.

\end{thebibliography}

\end{document}